\documentclass[journal]{IEEEtran}
\ifCLASSINFOpdf
\else
\fi
\usepackage[acronym,nonumberlist,sort=use]{glossaries}
\usepackage{circuitikz,tikz,svg,booktabs}
\usepackage{siunitx} 
\usepackage{cleveref} 
\usepackage{tablefootnote,amsmath}
\usepackage{soul}
\usepackage{cite}
\crefformat{figure}{Fig.~#2#1#3} 
\Crefformat{figure}{Fig.~#2#1#3}
\crefformat{section}{Section~#2#1#3} 
\crefformat{appendix}{appendix~#2#1#3}

\Crefformat{appendix}{Appendix~#2#1#3}

\newacronym{soc}{SoC}{System-on-Chip}
\newacronym{adc}{ADC}{Analog-to-Digital Converter}
\newacronym{ina}{INA}{Instrumentation Amplifier}
\newacronym{afe}{AFE}{Analog Front End}
\newacronym{1/f}{1/f}{Flicker noise}
\newacronym{fc}{F\textsubscript{c}}{flicker noise corner frequency }
\newacronym{fch}{$F_{ch}$}{Chopping frequency}
\newacronym{fin}{F\textsubscript{in}}{Input signal frequency}
\newacronym{am}{AM}{Amplitude modulation}
\newacronym{ccia}{CCIA}{Capacitively Coupled Instrumentation Amplifier}
\newacronym{opamp/ota}{OPAMP/OTA}{Operational amplifier/Operational transconductor amplifier}
\newacronym{cmfb}{CMFB}{Common Mode Feedback}
\newacronym{fft}{FFT}{Fast Fourier Transform}
\newacronym{llm}{LLM}{Large Language Models}
\newacronym{cmrr}{CMRR}{Common Mode Rejection Ratio}
\newacronym{sm}{SM}{Switch Matrix}
\newacronym{ugb}{UGB}{Unity Gain Bandwidth}
\newacronym{rtl}{RTL}{Register Transfer Level}
\newacronym{pnr}{PnR}{Place and Route}
\newacronym{id}{I\textsubscript{d}}{Drain current}
\newacronym{qlora}{QLoRA}{Quantized Low Rank Adaptation}
\newacronym{lora}{LoRA}{Low Rank Adaptation}
\newacronym{spice}{SPICE}{Simulation Program with Integrated Circuit Emphasis}
\newacronym{Zin}{$Z_{in}$}{Input Impedance}
\newacronym{Zos}{$Z_{o}$}{Sensor Output Impedance}
\newacronym{cpf}{$C_{pf}$}{Positive Feedback Capacitor}
\newacronym{ci}{$C_i$}{Input Capacitance}
\newacronym{cf}{$C_{f}$}{Feedback Capacitor}
\newacronym{sgfet}{SGFET}{Suspended Gate Field Effect Transistors}
\newacronym{mems}{MEMS}{Micro-Electro-Mechanical Systems}
\newacronym{dsl}{DSL}{DC-Servo-Loop}
\newacronym{ecg}{ECG}{Electrocardiogram}
\newacronym{snr}{SNR}{Signal to noise ratio}
\begin{document}
%
\title{A High Input Impedance Chopper Stabilized Amplifier Based On Charge Conservation}
%
%
%

\author{Prabhas~K~Deshpande\thanks{Prabhas K Deshpande and Naveen Kadayinti are with the department of EECE at IIT Dharwad, Karnataka, India - 580011 (e-mail: ee24ms002@iitdh.ac.in).}, 
        Naveen~Kadayinti\thanks{An Indian Patent Application No. 202641027947 based on
part of the work reported in this article has been filed in 10th March 2026.}}

%
%

\markboth{Electronics Systems Laboratory, IIT Dharwad}%
{Shell \MakeLowercase{\textit{et al.}}: A High Input Impedance Chopper Stabilized Amplifier Based On Charge Conservation}
%



\maketitle

\begin{abstract}
Chopper stabilized amplifiers are popularly used for realizing amplifiers with low offset and  for rejecting flicker noise. One of the main limitations of these amplifiers is the low \gls{Zin} produced by the switch capacitor input network. \gls{Zin} here is resistive due to the  switch capacitor action and is inversely proportional to the product of \gls{fch} and \gls{ci}. Since \gls{fch} should be  greater than the flicker noise corner frequency, this results in a low \gls{Zin}. When interfacing sensors with high \gls{Zos}, chopper stabilized amplifiers load the sensors resulting in reduced sensitivity. This paper presents a novel input impedance boosting technique -- \textit{Differential capacitor flipping technique} for chopper based \gls{ccia}, which prevents discharge and recharge of \gls{ci}'s in every cycle by reconfiguring the capacitor positions while preserving the chopping operation. This ideally results in a purely capacitive \gls{Zin} which is independent of \gls{fch}. The proposed architecture is used to demonstrate \gls{ecg} signal acquisition with dry electrodes that have \gls{Zos} in the order of a few Mega Ohms. This circuit implemented in TSMC 65~nm CMOS technology node  features \gls{Zin} of 21~G$\Omega$ at DC. The circuit has a power consumption of 2.6~$\mu$W (2.8~$\mu$W including clock generation circuits), with 7.2~$\mu V_{rms}$ (1~Hz-150~Hz) of total integrated input referred noise.
\end{abstract}


%
\IEEEpeerreviewmaketitle

\section{Introduction}
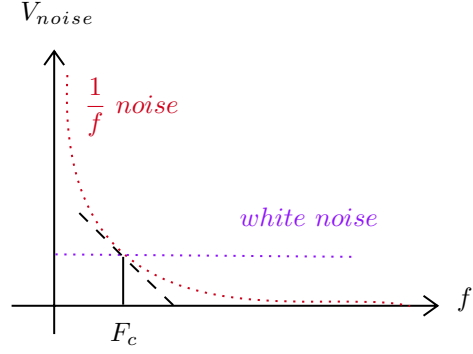
\begin{figure}
        \centering
		\tikzset{every picture/.style={line width=0.75pt}} 
			
			\begin{tikzpicture}[x=0.75pt,y=0.75pt,yscale=-1,xscale=1]
			
			\draw  (50,184.8) -- (263.75,184.8)(71.38,57) -- (71.38,199) (256.75,179.8) -- (263.75,184.8) -- (256.75,189.8) (66.38,64) -- (71.38,57) -- (76.38,64)  ;
			\draw [color={rgb, 255:red, 208; green, 2; blue, 27 }  ,draw opacity=1 ] [dash pattern={on 0.84pt off 2.51pt}]  (78,69) .. controls (71,215) and (206,174) .. (250,185) ;
			\draw [color={rgb, 255:red, 144; green, 19; blue, 254 }  ,draw opacity=1 ] [dash pattern={on 0.84pt off 2.51pt}]  (72,159) -- (223,160.25) ;
			\draw  [dash pattern={on 4.5pt off 4.5pt}]  (84,138.25) -- (106,160.25) -- (134,187.25) ;
			\draw    (106,160.25) -- (106,184.25) ;
			
			\draw (98,192) node [anchor=north west][inner sep=0.75pt]   [align=left] {$\displaystyle F_{c}$};
			\draw (163,134) node [anchor=north west][inner sep=0.75pt]  [color={rgb, 255:red, 144; green, 19; blue, 254 }  ,opacity=1 ] [align=left] {$\displaystyle {\textstyle white\ noise}$};
			\draw (85,69) node [anchor=north west][inner sep=0.75pt]  [color={rgb, 255:red, 208; green, 2; blue, 27 }  ,opacity=1 ] [align=left] {$\displaystyle \frac{1}{f} \ noise$};
			\draw (53,31) node [anchor=north west][inner sep=0.75pt]   [align=left] {$\displaystyle V_{n}{}_{o}{}_{i}{}_{s}{}_{e}$};
			\draw (272,174) node [anchor=north west][inner sep=0.75pt]   [align=left] {$\displaystyle f$};
			\end{tikzpicture}

		\caption{Flicker noise profile of \gls{ina}.}
		\label{fig: 1/f}
	\end{figure}

\begin{figure}[h!]
		\centering
		\includegraphics[scale=0.6]{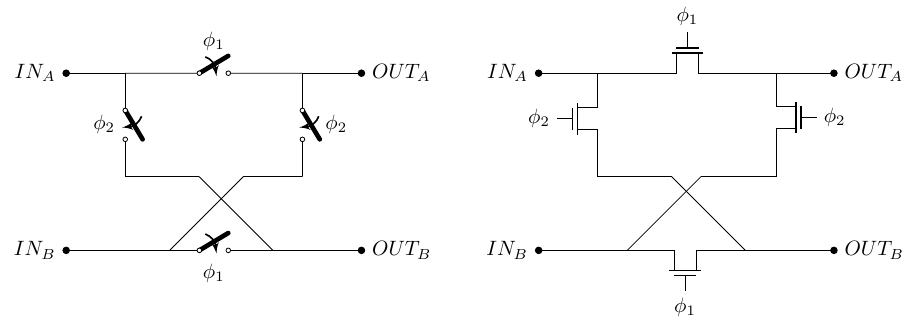}
		\caption{Chopper circuit: Ideal and Practical.}
		\label{fig: chopper}
	\end{figure}

%
%
%
%
\label{sec: introduction}
    \IEEEPARstart{A} key consideration when designing an \gls{ina} is its noise profile. Typically, there are two categories of noise that are predominant, the \gls{1/f} and the white noise. While white noise is omnipresent, \gls{1/f} noise dominates the lower end of the frequency spectrum. The flicker noise spectral density is given by

	\begin{equation}
		V_{noise} = \frac{K}{C_{ox}WL*f} \hspace{3mm} \frac{V}{\sqrt{Hz}}
		\label{eq: 1/f}
	\end{equation} 
	
	where `K' is the technology dependent constant, `C\textsubscript{ox}' is the gate oxide capacitance per unit area, `WL' is the  area of the transistor's channel and `f' is the frequency~\cite{razavi2005design}. The \gls{fc} is defined as the frequency below which \gls{1/f} noise dominates and above which thermal noise determines the noise power, as shown in ~\cref{fig: 1/f}.
	For sensors used to detect low frequency signals, for example bio potential signals, \gls{1/f} noise will dominate resulting in low \gls{snr}. One of the ways of improving the \gls{snr} is by sizing the transistors appropriately. Typically in amplifier circuits the input transistors contribute most to the 
	flicker noise. Hence, the size of these input transistors is usually kept very high to improve \gls{snr},
	\cref{eq: 1/f}. However, practical limitations restrict \gls{fc} to a few thousands of hertz~\cite{flick_ref}. Another reported way of mitigating the effect of \gls{1/f} noise is to upconvert the signal to a frequency beyond \gls{fc}. This is done by multiplying the input signal with a square wave, also known as chopping. Choppers are the blocks used to perform this modulation. Typical chopper realization is shown in \cref{fig: chopper}.
    
\begin{figure}
        \centering
        \includegraphics[scale=0.75]{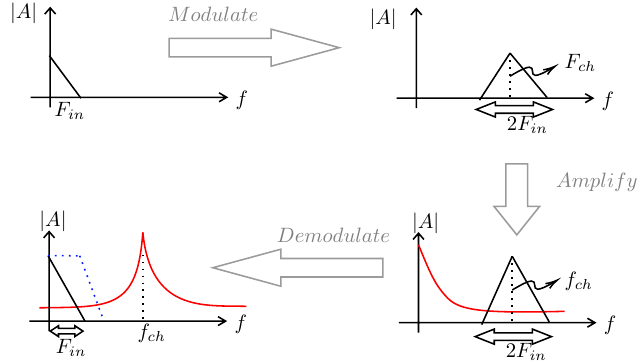}
        \caption{Chopper based modulation.}
        \label{fig: modulation}
    \end{figure}

    After upconverting, the signal is amplified and then downconverted. Once the signal is demodulated, the signals of interest comes back to base-band while flicker noise and offset get upconverted. This is pictorially shown in  \cref{fig: modulation}. 

    
	When a chopper is used as a preceding block to \gls{ccia} and terminated with another chopper for demodulation, the total input impedance of the chopper based \gls{ccia} will resemble a Switch Capacitor Resistor; \cref{eq: Zin_chop}. 
    
    \begin{equation}
    Z_{in} = \frac{1}{2F_{ch}C_i}
    \label{eq: Zin_chop}
    \end{equation}
    
    The periodic charging and discharging of \gls{ci} results in such an expression. For typical values of \gls{fch} and \gls{ci}, \gls{Zin} is in order of Mega Ohms, and this will be comparable to the output impedance of the sensors like Dry electrodes used for Bio potential application and pH sensors.
    
\begin{figure}[h!]
    \centering
    \includegraphics[width=\columnwidth]{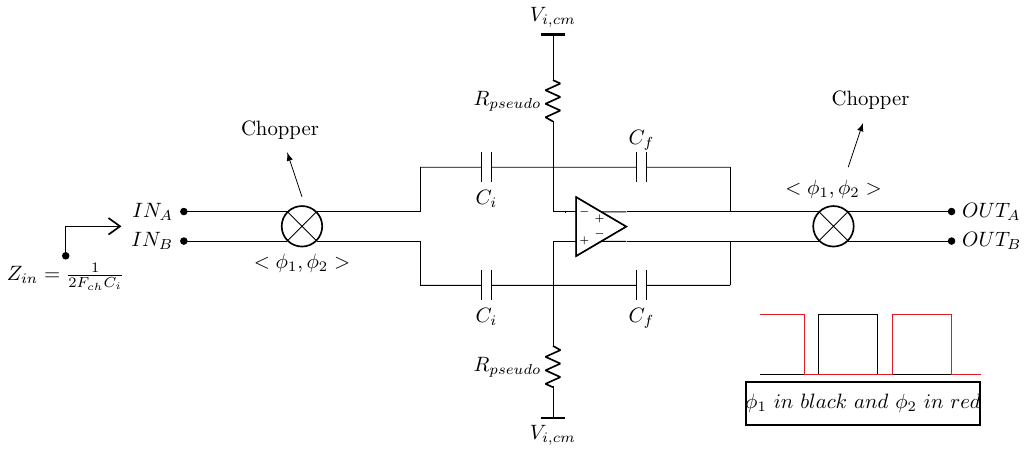}
    \caption{Typical CCIA with chopper. The chopper circuit is shown in \cref{fig: chopper}}
    \label{fig: CCIA_chopp}
	\end{figure}

	The complete system of \gls{ccia} with chopper is shown \cref{fig: CCIA_chopp}. Pseudo resistors; essentially off transistors are used to bias the input transistors of \gls{opamp/ota}.

\section{prior art for impedance boosting in \gls{ccia}}
\label{sec: prior_art}
As mentioned in \cref{sec: introduction} the \gls{Zin} of \gls{ccia} is comparable to \gls{Zos}. The signal strength available at the input of \gls{ccia} is given by \cref{eq: volt_div}.

\begin{equation}
    \beta = \frac{Z_{in}}{Z_{in} + Z_{o}} \\
    \label{eq: volt_div}
\end{equation}

Substituting \gls{Zin} from \cref{eq: Zin_chop} into \cref{eq: volt_div},

\begin{equation}
    \beta = \frac{1}{1+(2F_{ch}C_i)(Z_{o})}.
    \label{eq: volt_div_modi}
\end{equation}

The lower bound on \gls{fch} is ($\gls{fc} + \glsunset{fin}\gls{fin}/2$). This puts constraints on the choice of \gls{fch} that limits the highest value of $\beta$ that can be achieved. To remove the dependency of \gls{fch} from \cref{eq: volt_div_modi}, existing techniques boost the input impedance to high values. As a result of this under limiting condition of \gls{Zin}~$\to~\infty$ the value of \cref{eq: volt_div} is unity, independent of \gls{fch}. The two popular kind of \gls{Zin} boosting schemes are discussed below.

\subsection{Positive Feedback Based}
\label{subsec: pos_fdb}
\begin{figure}[h!]
			\centering
		      \includegraphics[scale=0.7]{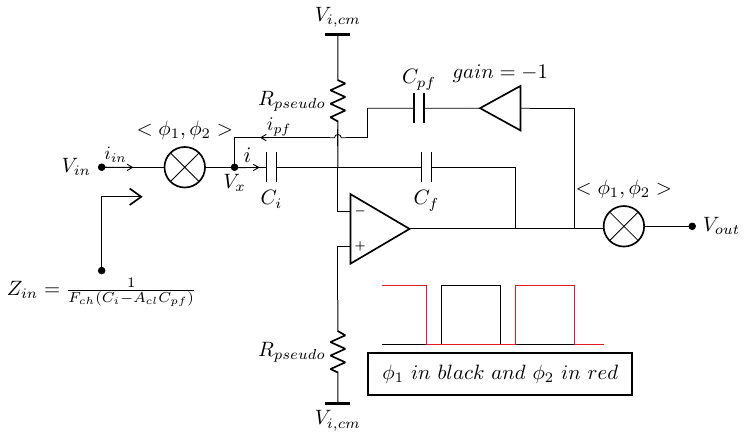}
			\caption{Single ended version of chopper based \gls{ccia} with positive feedback; ~\cite{xu2011160} reports $\ge 1G \Omega$ of impedance at DC.}
			\label{fig: CCIA_chopp_pos_fdb}
		\end{figure}

        The most widely reported impedance boost technique is the positive feedback based technique. From first principles, to boost \gls{Zin} the input current should be made zero or negligible. Due to the resemblance of input network with switch capacitor resistor kind of arrangement there is an average input current that flows through \gls{ci} of magnitude,
        \begin{eqnarray}
            i = v_{in}/Z_{in} \\
            i = v_{in}*\left(F_{ch} C_i \right) .
            \label{eq: in_current}
        \end{eqnarray}
        To nullify this, an equal and opposite current needs to be fed into \gls{ci} from some other path apart from the input source. Since the current is ought to be of opposite magnitude, positive feedback should be employed as shown in \cref{fig: CCIA_chopp_pos_fdb}. The required positive feedback current can be shown to be

        \begin{eqnarray}
            i_{pf} = v_o * \left(F_{ch} C_{pf}\right).
            \label{eq: pf_current}
        \end{eqnarray}

        Equating \cref{eq: in_current} and \cref{eq: pf_current}, we can write
        
        \begin{eqnarray}
            C_{pf} = \frac{C_i}{A_{cl}}
        \end{eqnarray} 
        where $A_{cl}$ is the closed loop gain of \gls{ccia}. Now, writing KCL at the node $V_x$: $i_{in} + i_{pf} = i$, since $i_{pf}$ and $i$ are equal, $i_{in} = 0$; $Z_{in} \to \infty$.

        This circuit can also be analyzed using Miller's theorem. Again consider  \cref{fig: CCIA_chopp_pos_fdb} by the virtue of Miller's effect the node $V_{x}$, will have a total capacitance of~$\approx~(C_{i}-A_{cl}C_{pf})$, 
        and hence according to \cref{eq: Zin_chop} the input impedance would be 

        \begin{equation}
                  Z_{in} =  \frac{1}{F_{ch}(C_i-A_{cl}C_{pf})}.  
                  \label{eq: zin_pos_fdb}
        \end{equation}

        If we manage to make $(C_i-A_{cl}C_{pf})=0$, infinite input impedance could be achieved~\cite{fan2013capacitively}. Reported works using this technique report $>$1~G$\Omega$ of \gls{Zin}~\cite{pos_ccia_2, china1, pos_ccia_3}. The parasitics at bond pads degrades developed \gls{Zin}, and this can be addressed by adding dual positive feedback loops as reported in~\cite{pos_ccia_4}. One to cancel \gls{ci} as discussed above and the other to cancel the parasitics at the bond pads. 
        
        Since the closed loop gain of any amplifier used in main path will be high, the value of $C_{pf}$ will turn out to be very small. From \cref{eq: zin_pos_fdb} it is evident that any mismatches in \gls{cpf} will drastically reduce \gls{Zin}. There are reported ways of solving this problem, course-fine tuning being a popular one as described in \cite{xu2011160}. Here, the value $C_{pf}$ could be calibrated to obtain the maximum input impedance based on digital logic. Alternatively varactors are also used to achieve the same thing~\cite{yan2024analog}.

\subsection{Auxiliary Path Based}
\label{subsec: aux_fdb}
		Another widely used technique is the precharge-based approach. Like positive feedback method, it aims to nullify the current drawn from the input source — however, instead of a positive feedback loop, it employs a feedforward or auxiliary path. This auxiliary path is selectively activated so that the average input current, which would otherwise be sourced from the input, is instead supplied through the auxiliary path. 
        
        Fundamentally, this works on the fact that the entire circuit described in  \cref{fig: CCIA_chopp} is halted 
		for some time between the non-overlapping clocks of choppers ($\phi_1$ and $\phi_2$). Meanwhile \gls{ci} is charged to the upcoming value of the input signal through the auxiliary path. When the circuit resumes back the input source doesn't need to supply any charge to \gls{ci} as it is already charged through the auxiliary path, and hence \gls{Zin} is boosted. 
        
\begin{figure}[h!]
\tikzset{every picture/.style={line width=0.75pt}} 

\begin{tikzpicture}[x=0.75pt,y=0.75pt,yscale=-0.7,xscale=0.7]

\draw (100,132) -- (450,122);

\foreach \x in {115, 121, ..., 439} {
    \draw [color={rgb, 255:red, 208; green, 2; blue, 27 }, draw opacity=1] 
        (\x,181) -- (\x+1,181) -- (\x+1,141) -- (\x+4,141) -- (\x+4,181) -- (\x+6,181);
}

\draw [color={rgb, 255:red, 208; green, 2; blue, 27 }, dashed] (116,141) -- (116,131.5);
\draw [color={rgb, 255:red, 208; green, 2; blue, 27 }, dashed] (119,141) -- (119,131.4);

\draw [<-] (117.5,129) .. controls (105,95) and (85,90) .. (102.84,75.88) ;
\draw (66,55) node [anchor=north west][inner sep=0.75pt] [align=left] {$\text{almost constant}$};

\end{tikzpicture}
\caption{Signal characteristics during chopping.}
\label{fig: sig_char}
\end{figure}
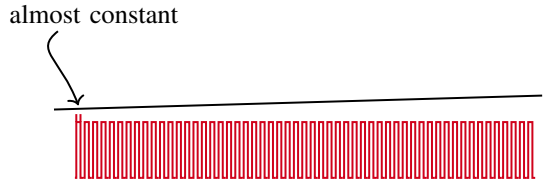

        The input signal being a slow moving one, compared to chopping signal facilitates such precharge operation (see \cref{fig: sig_char}). To prevent aliasing, \gls{fch} should be at least twice \gls{fin}. This relation is overshadowed by the fact that \gls{fch} also needs to be greater than \gls{fc} which is usually larger than bandwidth of input signal.
        
\begin{figure}[h!]
			\centering
			\includegraphics[scale=0.5]{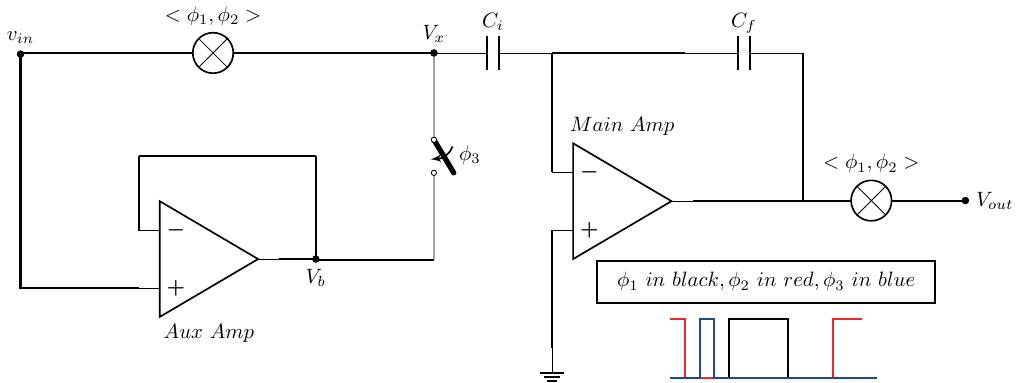}
			\caption{Single ended version of precharge based \gls{ccia} with chopper; ~\cite{chandrakumar2017high} reports $300~M\Omega$ of impedance at DC. }
			\label{fig: CCIA_chopp_aux}
		\end{figure}

        The circuit in  \cref{fig: CCIA_chopp_aux} described in~\cite{chandrakumar2017high} is the typical way of implementing such a precharge operation. Here, the clocks $\phi_1 $ and $\phi_2$ serve their usual purpose as in a normal chopper based \gls{ccia}, but the clock $\phi_3$ being orthogonal to both $\phi_1 \ $ and $\ \phi_2$ turns on the precharge path just before $\phi_1$ turns on and charges \gls{ci}.
        
        Now, ideally the Aux amp in  \cref{fig: CCIA_chopp_aux} would charge \gls{ci} to $v_{in}$, but the finite bandwidth~($\omega=\tau^{-1}$) and the gain error~($\alpha$) of the Aux amp will charge the node $V_{x}$ to $v_{in}$ asymptotically. Hence, node $V_x$ will be charged to $v_{in}* ( \alpha - \exp{\frac{-T}{\tau})}$, where $T$ is the time available for the precharge path. The difference between the actual and precharged value (denoted by $\Delta v$), will be equal to $ v_{in} * ( \alpha + \exp\frac{-T}{\tau}) $. 
        
        \gls{Zin} can be derived as follows. The residual charge that needs to be built upon \gls{ci} by input source is,
        \begin{align}
            Q = C_i* \Delta v  \\
            Q  = C_i* v_{in} *(  \alpha + \exp\frac{-T}{\tau})  .
        \end{align}
            Multiplying by \gls{fch} on both sides, 
        \begin{align}
            Q \cdot F_{ch} &= C_i \cdot v_{in} \cdot \left( \alpha + \exp\left( \frac{-T}{\tau} \right) \right) \cdot F_{ch} \\
            I &= C_i \cdot v_{in} \cdot \left( \alpha + \exp\left( \frac{-T}{\tau} \right) \right) \cdot F_{ch} \\
            \frac{v_{in}}{I} &= \frac{1}{C_i \cdot \left( \alpha + \exp\left( \frac{-T}{\tau} \right) \right) \cdot F_{ch}} \\
            Z_{in} &= \frac{\exp\left( \frac{T}{\tau} \right)}{F_{ch} \cdot C_i \cdot \left( \alpha \exp\left( \frac{T}{\tau} \right) + 1 \right)} \\
            Z_{in} &= Z_o \cdot \frac{\exp\left( \frac{T}{\tau} \right)}{\alpha \exp\left( \frac{T}{\tau} \right) + 1} 
            \label{eq: Zin_CCIA_chopp}
        \end{align}
        where $Z_o$ is the input impedance of a regular chopper based \gls{ccia} described in \cref{eq: Zin_chop}\footnote{Factor 2 is absent as the derivation is performed for single ended version.}. Under limiting conditions of $\tau~\to~0$ and $\alpha~\to~0$, \gls{Zin}~$\to~\infty$.

        This technique has a few limitations. The \gls{1/f} noise and offset of the axillary path will remain unchopped in the signal chain as there are no choppers in this path. This contributes to increased input referred noise apart from extra power used at auxiliary amplifiers. The authors of~\cite{aux_2} use a chopper stabilized auxiliary amplifier to mitigate above discussed issues. Since $\phi_3$ is selectively turned on in the dead zone between $\phi_1$ and $\phi_2$, chopping of \gls{1/f} and offset of auxiliary amplifier should happen in this tiny interval of time. 

\subsection{Use of \gls{dsl}}
\begin{figure}
    \centering
    \includegraphics[width=\columnwidth]{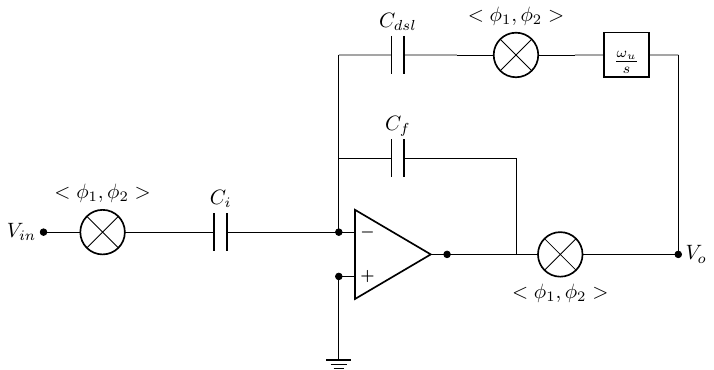}
    \caption{\gls{ccia} with \gls{dsl} at block level. Here $\omega_u$ is the unity gain frequency of the integrator used.}
    \label{fig: dsl_block}
\end{figure}

In many of sensor instrumentation applications a high pass corner is necessary to remove artifacts from the input signal. For example, a high pass corner at 1~Hz is usually used in Bio potential signal acquisition to remove motion artifacts and electrode offsets. Typically \gls{dsl} is used to achieve this, which consumes additional power and area. \cref{fig: dsl_block} shows block level realization of typical \gls{ccia} with \gls{dsl}. 

To derive the transfer function this block, assume ideal \gls{opamp/ota}. Writing KCL at inverting terminal,

\begin{align}
sC_i \cdot V_i &= sC_f \cdot V_o + sC_{dsl} \cdot\left(\frac{\omega_u \cdot V_o}{s}\right) \\
\frac{V_o}{V_i} &= \frac{C_i}{C_f + \frac{\omega_u C_{dsl}}{s}}
\label{eq: dsl}
\end{align}

From \cref{eq: dsl}, the midband gain is the  same as in the case of typical \gls{ccia}. In addition, there is a high pass corner at

\begin{equation}
    \omega_{h} = \frac{\omega_u* C_{dsl}}{C_{f}}.
    \label{eq: hpc_dsl}
\end{equation}

This high pass corner should have a value less than 1~Hz. The impedance boosting techniques mentioned in \cref{sec: prior_art} incorporate \gls{dsl} loop to create such a high pass corner.


%

\section{Differential Capacitor Flipping Technique for \gls{Zin} Boosting}
\label{sec: novel_arch}
In this section, a novel architecture is proposed to boost the \gls{Zin} of chopper based \gls{ccia}. This method takes advantage of the stored charges on the differential capacitors to boost \gls{Zin} by periodically reconfiguring them. In addition, it also addresses the issues of existing impedance boosting techniques, namely strict matching constraints, additional auxiliary circuit that contributes noise and the need for \gls{dsl}.

\subsection{Key Observation: Charge symmetry in \gls{ccia}}
\begin{figure}[h!]
			\centering
			\includegraphics[scale=0.7]{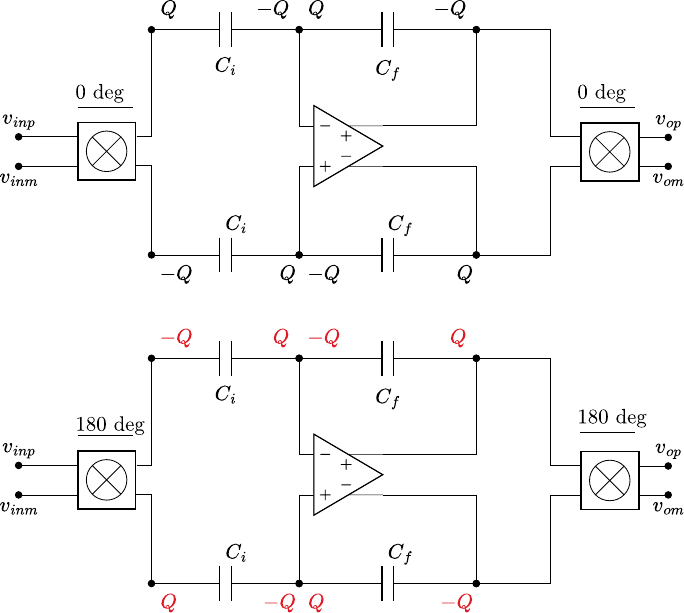}
			\caption{Charge pattern in typical \gls{ccia}.}
			\label{fig: chg_patt}
		\end{figure}

        A careful look at \gls{ccia} shown in  \cref{fig: chg_patt} reveals an interesting charge symmetry between the differential capacitors in two phases of the chopping cycle. The charges on differential capacitors, both input and feedback, essentially have same magnitude but opposite signs, which also seems obvious due to the presence of chopper block which flips the inputs periodically. Therefore, a meticulous rearrangement of these capacitors can harness the existing charge stored on differential capacitors.
        
        As discussed earlier in \cref{sec: prior_art}, to prevent aliasing \gls{fch} should be at least twice the \gls{fin}. Since, \gls{fc} will be higher than \gls{fin} no information is lost by modulating the input signal - this is shown in \cref{fig: sig_char}.

		\subsubsection{Differential capacitor flipping}
		The above discussed charge symmetry feature could be leveraged to get high input impedance by flipping the differential capacitors both input and feedback. This can be done when the circuit just enters either of the chopping phases. Doing so will guarantee that the input source does not  need to supply the required charges in a given chopping phase, as the same amount of it is already available on the other differential side which is now flipped. This in principle means: infinite input impedance, \gls{Zin}~$\to~\infty$.

    	\subsubsection{How to flip?}
\begin{figure}[h!]
			\centering
			\includegraphics[scale=1]{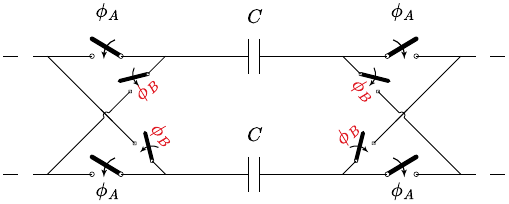}
			\caption{Choppers can be used to flip the differential capacitors.}
			\label{fig: how2flip}
		\end{figure}

		To perform flipping of differential capacitors we need a switching circuit as described in  \cref{fig: how2flip}. $\phi_A$ and $\phi_B$ are non-overlapping clocks, when $\phi_A$ is turned on, the position of capacitors is intact and when $\phi_B$ is turned on, the capacitors are flipped. Coincidentally this again resembles to using choppers where the differential capacitors that are supposed to be flipped are sandwiched in between two choppers. 




\subsection{Proposed Circuit: Leveraging stored charges to boost \gls{Zin}}
\label{subsec: proposed_arch}
\begin{figure*}[t]
    		\centering
    		\includegraphics[width=\textwidth]{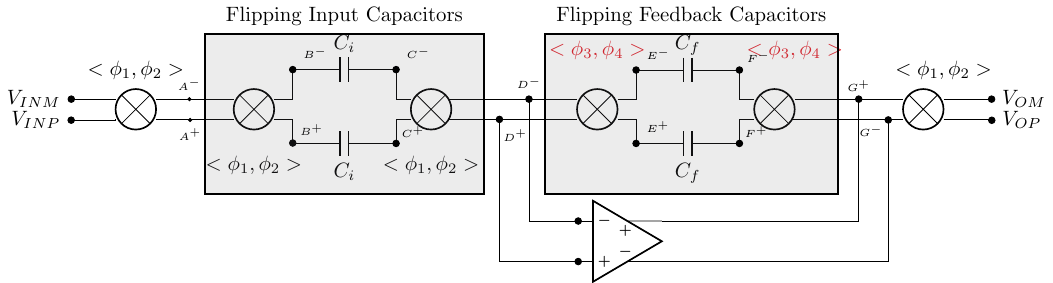}
			\caption{Final architecture at block level.}
			\label{fig: fin_arch}
	\end{figure*}

        Incorporating this method into typical \gls{ccia} results in the circuit shown in \cref{fig: fin_arch}, where $\phi_1  \ \& \ \phi_2$ are used for modulation and flipping input differential capacitors; $\phi_3 \ \& \ \phi_4$ used to flip feedback differential capacitors. Interestingly, this architecture has an implicit high pass corner whose location depends on the order of flipping  of the input and feedback capacitors. This point will be elaborated in \cref{subsec: tf_fin_arch}. 

    \begin{figure}[htbp]
        \includegraphics[width=\columnwidth]{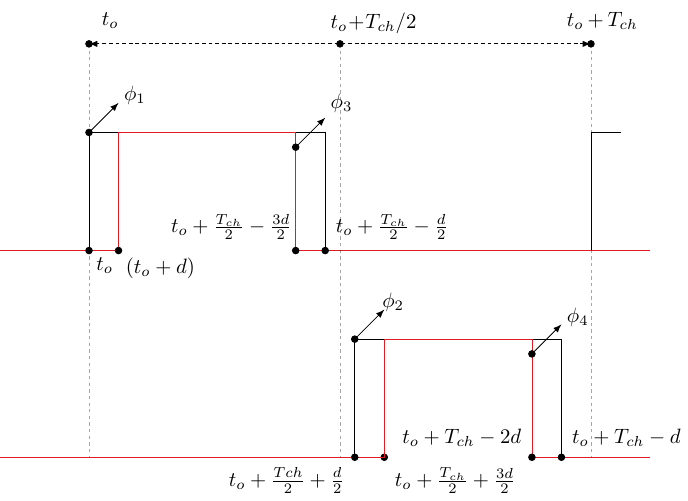}
        \caption{Two sets of non-overlapping clocks.}
        \label{fig: clks_seq} 
    \end{figure}

\subsubsection{Deriving the input impedance}
	From  \cref{fig: clks_seq} and \cref{fig: fin_arch}, in a given chopping cycle, the following cycle of operations follows.
    \begin{itemize}
        \item In the first half of chopping phase,
            \begin{enumerate}
                \item At $t_o$, the \gls{ci}'s are connected as usual to nodes: $(A^+,A^-)$ and $(D^+,D^-)$ .
                \item At $(t_o + d)$, the \gls{cf}'s are connected as usual to nodes: $(D^+,D^-)$ and $(G^+,G^-)$.
                \item At $(t_o + T_{ch}/2- 3d/2)$, \gls{cf}'s get completely disconnected.
                \item At $(t_o + T_{ch}/{2}- {d}/{2})$, \gls{ci}'s get completely disconnected. Hence, the last value sampled on the \gls{ci}'s will be $v_{in}\left(t_o + \frac{Tch}{2}-\frac{d}{2}\right)$.
            \end{enumerate}
        \item In the other half of chopping phase,
            \begin{enumerate}
                \item At $(t_o + {T_{ch}}/{2} + {d}/{2})$, \gls{ci}'s are flipped and connected to nodes: $(A^+,A^-)$ and $(D^+,D^-)$.
                \item At $(t_o + T_{ch}/{2} + {3d}/{2})$, \gls{cf}'s are flipped and connected to nodes: $(D^+,D^-)$ and $(G^+,G^-)$.
                \item At time $(t_o + T_{ch} - 2d)$ and $(t_o + T_{ch} -d)$, \gls{cf}'s and \gls{ci}'s are disconnected respectively. Hence, \gls{ci}'s need to charge to a value of $v_{in}\left(t_o + T_{ch} -d\right)$ by the end of chopping interval.
            \end{enumerate}    

    \end{itemize}
    
   The average input current draw to charge \gls{ci}'s, in a given chopping period can be expressed as, 
	\begin{eqnarray}
		I_{avg} &=& \frac{\Delta Q}{T_{ch}} \\
		I_{avg} &=& 2C_{i}* \frac{\Delta v_{in}} {T_{ch}}
	\end{eqnarray}


\begin{equation}
\begin{split}
    I_{avg} &= \frac{2 \cdot C_{i} \cdot \Big[ v_{in}(t_o + T_{ch} - d) }{T_{ch}} \\
            &\quad \frac{- v_{in}\left(t_o + \frac{T_{ch}}{2} - \frac{d}{2}\right) \Big]}{T_{ch}}
\end{split}
\end{equation}

\begin{equation}
\begin{split}
    I_{avg} &= \frac{2 \cdot C_{i} \cdot \Big[ v_{in}(t_o + T_{ch} - d) }{2 \cdot \frac{T_{ch}}{2}} \\
            &\quad \frac{- v_{in}\left(t_o + \frac{T_{ch}}{2} - \frac{d}{2}\right) \Big]}{2 \cdot \frac{T_{ch}}{2}}
\end{split}
\end{equation}

   The non-overlapping delay $d$, is smaller in comparison with $T_{ch}$. Hence, it can be neglected for further calculations. The chopping period~($T_{ch}$) will be smaller than the input signal period, so, $I_{avg}$ can be approximated as

	\begin{equation}
		I_{avg} \approx C_{i}*\frac{dv_{in}}{dt} .
		\label{eq: fin_imp_2}
	\end{equation}

	This expression resembles that of current through a capacitor. Hence, this architecture provides a purely capacitive input impedance. \gls{Zin} can be written as,
    \begin{equation}
		Z_{in}= \frac{1}{\omega_{in} C_{i}}.
		\label{eq: fin_imp}
	\end{equation}


    From \cref{eq: fin_imp}, \gls{Zin} only depends on \gls{ci} and the input frequency. There is \textbf{no dependence of \gls{Zin} on chopping frequency (which is the major limitation of all existing architectures)}. The gain of this circuit is the same as in a typical \gls{ccia}: ${C_i}/{C_f}$. 
    
    To summarize, the behavior of this modified chopper based \gls{ccia} is exactly the same as that of a typical \gls{ccia} with added benefit of flicker noise mitigation. 
   

    



\subsubsection{Clock generation circuit}
\begin{figure}[h!]
    \centering
    \includegraphics[width=\columnwidth]{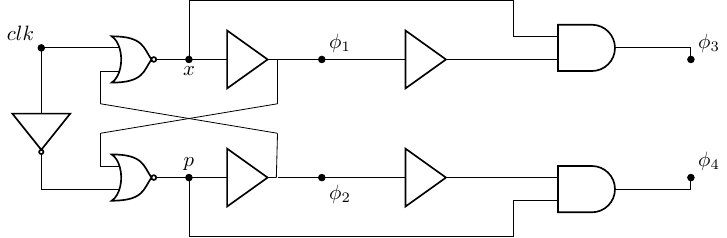}
    \caption{Clock generation circuit.}
    \label{fig: phi12}
\end{figure}

As shown in  \cref{fig: clks_seq} we need two sets of non-overlapping clocks whose relative timings are as indicated in~\cref{fig: clks_seq}. $\phi_1$ and $\phi_2$ can be generated from the conventional non-overlapping clock generation circuit shown in~\cref{fig: phi12}. The buffers used in this figure provide delay of `d' time units. The clock signals $\phi_3$ and $\phi_4$ can be generated from $\phi_1$ and $\phi_2$, as shown in~\cref{fig: phi12}.

\section{Effect of parasitics: The good and the bad}
\label{subsec: tf_fin_arch}

\cref{sec: novel_arch} presented the analysis of the proposed circuit with ideal components. This section explains the proposed circuit taking into account parasitic impedances.

	\begin{figure*}[t]
    		\centering
    		\includegraphics[width=\textwidth]{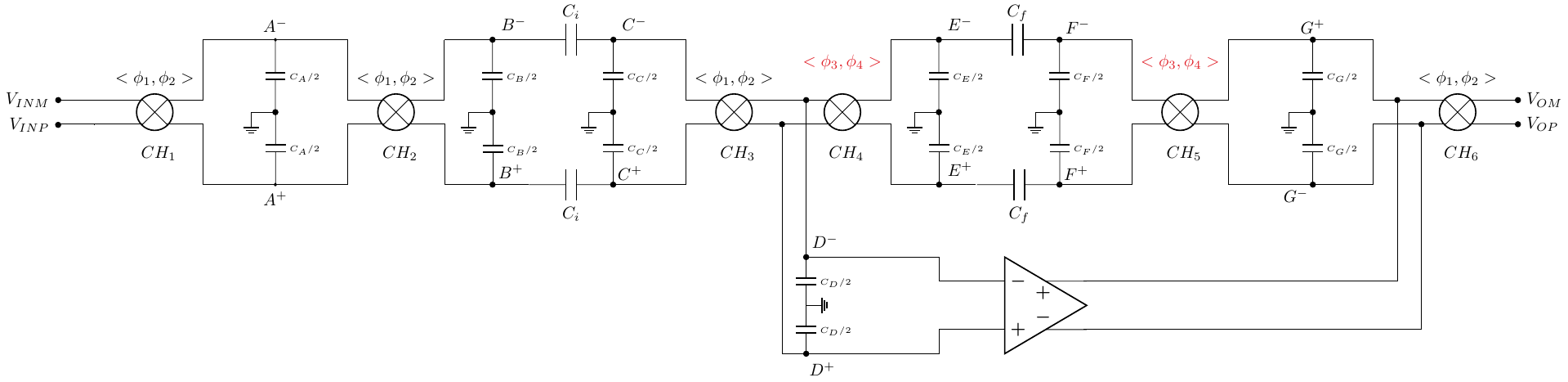}
			\caption{Final architecture at block level including parasitic capacitances.}
			\label{fig: fin_arch_new}
	\end{figure*}

\begin{figure}[h!]
    \centering
    \includegraphics[width=\columnwidth]{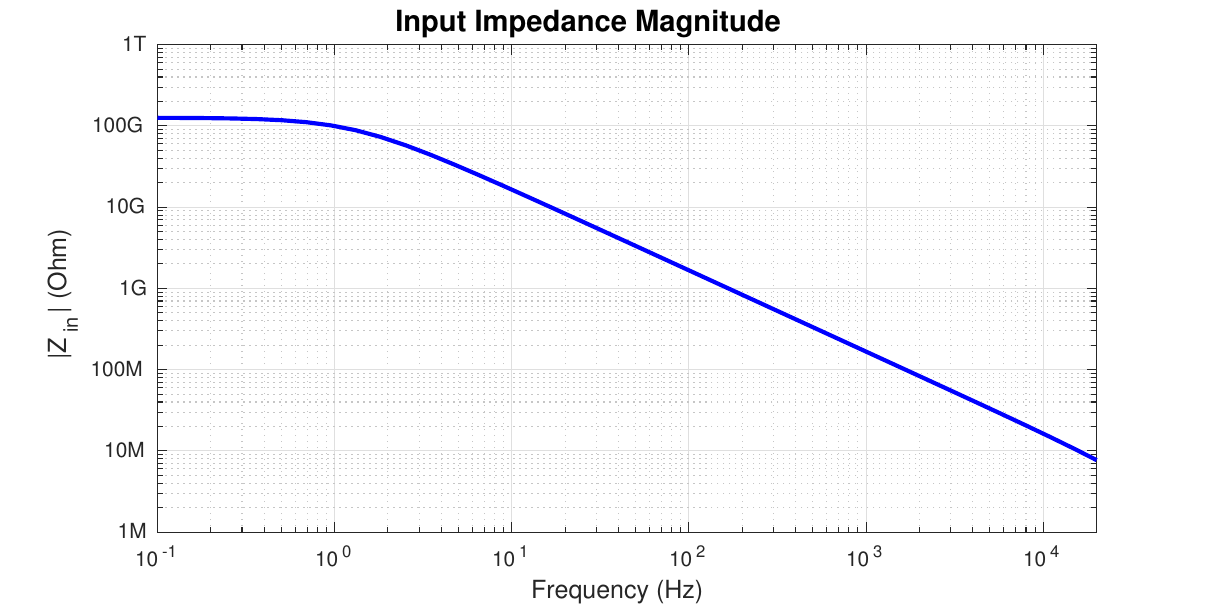}
    \caption{\gls{Zin} characteristics of proposed circuit.}
    \label{fig: zin_ideal}
\end{figure}

 A careful look at the nodes $(A^+,A^-)$ and $(B^+,B^-)$ in  \cref{fig: fin_arch_new} reveals that any parasitic capacitance between these nodes along with chopper $CH_{1}$ and $CH_2$ respectively, will manifest as a switch capacitor resistor. These resistor appear between $V_{INP}$ and $V_{INM}$ and the value of these resistors is determined by \cref{eq: Zin_chop}. Since the parasitic capacitance is usually small, the corresponding resistance value would be very high. Hence, their presence is inconsequential at other frequencies except at DC when \gls{ci}'s are open circuited. Hence, \gls{Zin} will have a  constant value near DC. As frequency increases it behaves as a capacitive impedance, as shown in \cref{fig: zin_ideal}. The \gls{Zin} relation can be expressed as,

	\begin{equation}
		Z_{in}= \frac{1}{2F_{ch}C_p} || \frac{1}{\omega_{in} C_{i}},
		\label{eq: fin_imp_new}
	\end{equation}
where $C_p$ is the equivalent parasitic capacitance.

\subsection{Transfer function of the proposed circuit}
\begin{figure}
    \centering
    \includegraphics[width=\columnwidth]{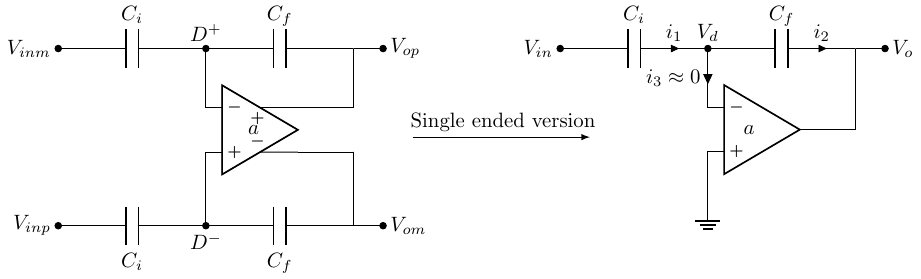}
    \caption{Typical \gls{ccia} and its single ended equivalent,}
    \label{fig: typ_ccia}
\end{figure}
Consider a typical \gls{ccia} shown in \cref{fig: typ_ccia}. For the purpose of analysis for deriving the transfer function we shall use the single ended version. Assuming non-ideal \gls{opamp/ota}, its transfer function can be derived as follows. Writing KCL at inverting terminal of \gls{opamp/ota}
\begin{align}
    (V_d & - V_i) \cdot sC_i + (V_d - V_o) \cdot sC_f = 0 \\
    V_o &= -aV_d \\
    V_d & = \frac{V_i \cdot C_i}{C_i + (1+a) \cdot C_f}.
\end{align}

Solving for $\frac{V_o}{V_i}$,
\begin{equation}
    \frac{V_o}{V_i}  = -\frac{a \cdot C_i}{C_i + (1+a) \cdot C_f}.
\end{equation}

As $a$ $\to~\infty$, 

\begin{equation}
    \frac{V_o}{V_i} = \frac{C_i}{C_f}.
    \label{eq: ccia_tf}
\end{equation}

In case of a chopper based \gls{ccia} shown in~\cref{fig: CCIA_chopp} the gain is given by ratio of feedback impedance to input impedance, which is indeed $\frac{C_i}{C_f}$.

\begin{figure}[h!]
    \centering
    \includegraphics[width=\columnwidth]{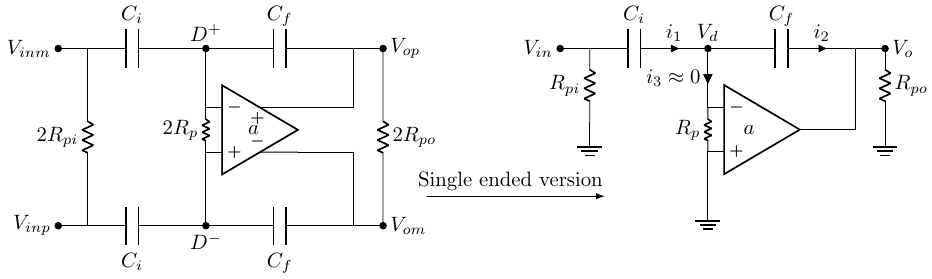}   
    \caption{Single ended version of proposed architecture in amplifying phase.}
    \label{fig: CCIA_App}
\end{figure}

In order to derive the transfer function for the proposed \textit{Differential Capacitor Flipping Technique}, we need to construct a small signal equivalent of the circuit in \cref{fig: fin_arch_new}. This is shown in~\cref{fig: CCIA_App}. The choppers $CH_1$ and $CH_2$ are responsible for the creation of a parasitic switch capacitor resistors $R_{A}$ and $R_{B}$. These can be clubbed as $R_{pi}$. Similarly, $R_{po}$ is composed of $R_G$ and $R_F$ that are created by the chopper $CH_5$. $R_{pi}$ loads the signal source resistively and $R_{po}$ loads \gls{opamp/ota} resistively. Both of these result in a gain reduction only. However, the choppers $CH_3$ and $CH_4$ also produce a similar resistor $R_p$ at the terminals of \gls{opamp/ota} which influences the transfer function which is discussed as follows.

To compute the transfer function, consider the single ended version of the circuit shown in~\cref{fig: CCIA_App}. Let  $V_d$ be the inverting terminal, $a$ be the open loop DC gain, and $R_{p}= \frac{1}{F_{ch}\widehat{C_p}}$ where $\widehat{C_p}$ is the equivalent parasitic capacitance at the inverting terminal. Solving for transfer function, we have, 

\begin{equation}
    V_o = -aV_d
    \label{eq: CCIA_test}
\end{equation}

Writing KCL at the inverting terminal of \gls{opamp/ota},
\begin{eqnarray}
    (V_d-V_{in})sC_i + \frac{V_d}{R_{p}}+(V_d+aV_d)sC_f=0 \\ 
    V_d\{s(C_i+(1+a)C_f)+\frac{1}{R_{p}}\}=V_{in}sC_i
\end{eqnarray}
    from  the above  equation  and  \cref{eq: CCIA_test}
\begin{eqnarray}
    \frac{V_{o}}{V_{in}} = \frac{-s(aC_{i})}{\frac{1}{R_{p}}+sC_{i}+s(1+a)C_{f}}
    \label{eq: CCIA_eq_2}
\end{eqnarray}

The above transfer function has a high pass corner at 

\begin{align}
    \omega_h &= \frac{1}{R_p*C_i*(1+a)C_f} \\ 
    & = \frac{1}{(\frac{1}{F_{ch}\widehat{C_{p}}})*(C_i+(1+a)C_f)} \\
    & \approx \frac{F_{ch}\widehat{C_p}}{(C_i+aC_f)} .
    \label{eq: hpc}
\end{align}

This transfer function shows that the proposed circuit has a high-pass nature and hence, cannot be used for DC signals. Most bio-potential signals like ECG/EMG lie in 1~Hz to 500~Hz bandwidth and need to reject motion artifacts and electrode offsets which are signals of frequencies less than 1~Hz. \Cref{eq: hpc} can be used to design the amplifier to have a high-pass corner less than 1~Hz. It may be noted that conventional architectures of increasing \gls{Zin} use \gls{dsl} for realizing this high-pass corner, which comes with its own complexity and power dissipation. Although \cref{eq: CCIA_eq_2} looks like an ideal high pass filter, it is limited by the finite bandwidth of the \gls{opamp/ota} used. The key parameters of this architecture (for differential version) required for the design are summarized in~\cref{tab: char_CCIA}.

\begin{table}[h!]
\renewcommand{\arraystretch}{2.5}
\caption{Parameters of thus develop \gls{ccia}.}
\label{tab: char_CCIA} 
\centering
\begin{tabular}{|| c  | c ||}
\hline
Design parameter & Expression \\
\hline
Input Impedance  & $\frac{1}{\omega_{in} C_{i}}$ \\
\hline
Mid band Gain     & $\frac{C_i}{C_f}$     \\
\hline
High Pass Corner & $\frac{2F_{ch}\widehat{C_p}}{(C_i+aC_f)}$ \\  
\hline
\end{tabular}
 
\end{table}

The high pass corner located at $\omega_{h}$  is  susceptible to the switching sequence (i.e. the relative timing relation between $\phi_1,\phi_2,\phi_3~\&~\phi_4 $). Let us consider a few switching patterns.   

\begin{itemize}

    \item \textbf{Pattern 1:}\begin{quote} The input and output is disconnected (using $CH_1$ and $CH_6$) before any capacitors are flipped. Consider that the feedback capacitors are flipped first within each cycle (i.e., toggle $CH_4$ and $CH_5$). This causes  a small potential at nodes $(D^+,D^-)$ due to the charge sharing between reoriented $C_f$ and $C_D$. The magnitude of this potential will be $C_iV_{in}/aC_f$. When the input capacitors flip and connect, they
    undergo charge redistribution as shown in \cref{fig: sw_pat_01}. This redistribution causes periodic charge-discharge action of $C_{D}$ where the corresponding current actually comes from the input source and leads to a lower value of $R_p$. This pushes the high pass corner to a higher frequency.
    \end{quote}
    
    \begin{figure}[h!]
        \centering
        \includegraphics[width=\columnwidth]{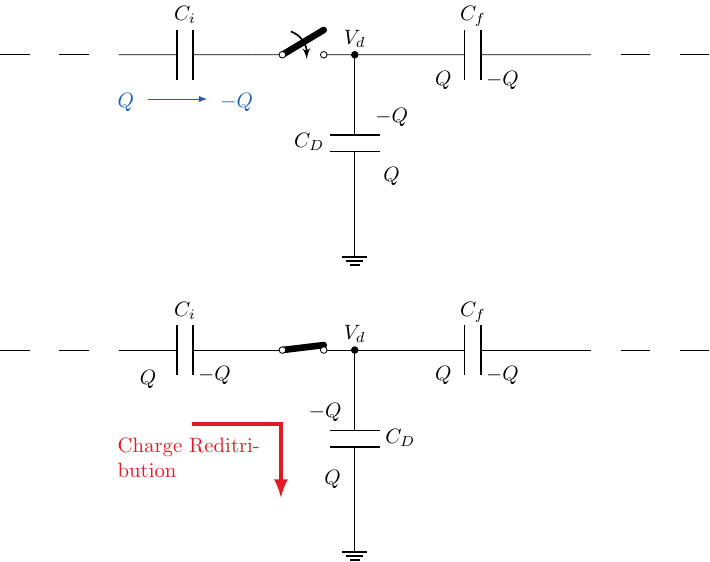}
        \caption{Charge redistribution: Input capacitors and parasitic capacitor at \gls{opamp/ota} terminals.}
        \label{fig: sw_pat_01}
    \end{figure}
    
    \item \textbf{Pattern 2:} \begin{quote} The input and output is disconnected (using $CH_1$ and $CH_6$) before any capacitors are flipped. Consider the  case when input capacitors are flipped first (i.e. toggle $CH_2$ and $CH_3$). This causes  a very small potential at nodes $(D^+,D^-)$ due to the charge sharing between reoriented $C_i$ and $C_D$. When the feedback capacitors flip and connect, they undergo charge redistribution as shown in \cref{fig: sw_pat_02}. Here, the charge-discharge current corresponding to $C_D$ essentially comes through $C_f$ from \gls{opamp/ota}. This produces better results in terms of proximity of the high pass corner towards the origin compared to the former pattern. However, a change in the node voltage of $V_d$ draws residual current from the input source, which may deviate the value of the high-pass corner from the analytical calculation. 
    \end{quote}
    
    \begin{figure}[h!]
        \centering
        \includegraphics[width=\columnwidth]{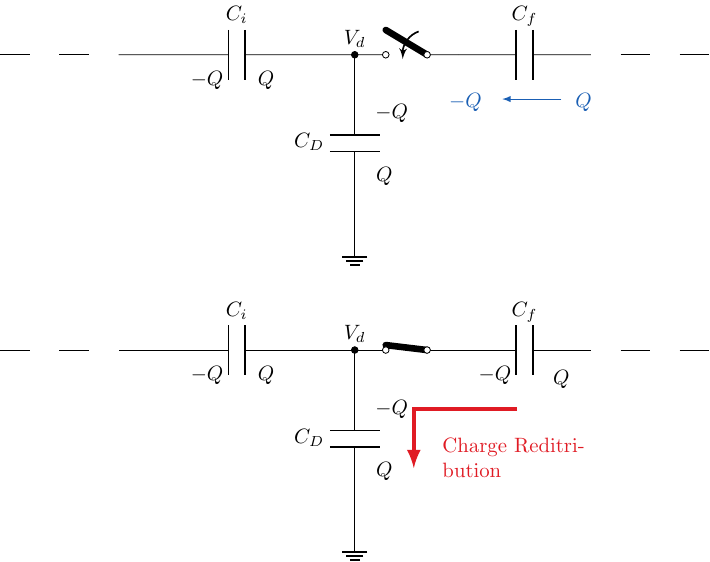}
        \caption{Charge redistribution: Feedback capacitors and parasitic capacitor at \gls{opamp/ota} terminals.}
        \label{fig: sw_pat_02}
    \end{figure}

\end{itemize}



\section{Example design: \gls{ecg} signal acquisition}
\label{sec: ecg}
This section shows an example design for \gls{ecg} signal acquisition using dry electrodes with an ROIC employing the proposed impedance boosting technique. Dry electrodes have a high value of \gls{Zos}, around 10M$\Omega$ at DC~\cite{dry_ele}. \gls{ecg} signals lie in the frequency spectrum of 1~Hz-150~Hz  ~\cite{zhu2022acquisition}. This ensures minimal clock feed-through effects. We also need to block DC to eliminate motion artifacts and electrode offsets. A high pass filter with corner frequency below 1~Hz is desired. 

An example design is implemented in TSMC~65~nm CMOS technology. The choice of design parameters is as follows.
\begin{itemize}

    \item Say we need at least \gls{Zin}=8*\gls{Zos}=80~M$\Omega$ at 150~Hz. Then from \cref{eq: fin_imp}, \gls{ci} $\approx$ 13~pF.

    \item Let the closed loop gain be 20~V/V. Then \gls{cf}=650~fF.

    \item From \cref{eq: CCIA_eq_2} having a high open loop gain (``a") helps in pulling the high pass corner closer towards the origin. Let \gls{opamp/ota} have a=60~dB.

    \item The value of \gls{fch} can be decided based on \gls{fc} of \gls{opamp/ota} designed. Hence, even the high pass corner can be decided once \gls{fch} is chosen.

    \item A load capacitance of 1~pF is used at the output nodes to model the subsequent stages.
\end{itemize}

\subsection{Folded cascode \gls{opamp/ota}}
\begin{figure}[h!]
    \centering
    \includegraphics[width=\columnwidth]{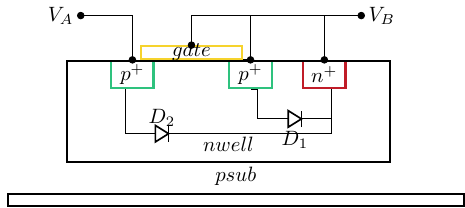}
    \caption{Pseudo resistor: Off transistor}
    \label{fig: off_Q}
\end{figure}

A folded cascode \gls{opamp/ota} is designed with a gain of 60~dB with \gls{ugb} of 1.5~MHz. Biasing of the input transistors was done using pseudo resistors~\cite{10509940}, which ensures that the parasitic diodes are at zero bias and hence, consumes no current ideally (refer \cref{fig: off_Q}). At the time of start-up, all nodes are discharged, V\textsubscript{B} $\approx$~0. Diodes D\textsubscript{1} and D\textsubscript{2} get forward biased and take V\textsubscript{B} to desired common mode level. The moment V\textsubscript{A}~=~V\textsubscript{B} diodes get into zero bias condition and provide huge resistance.

\begin{figure*}[t]
    \centering
    \includegraphics[scale=0.8]{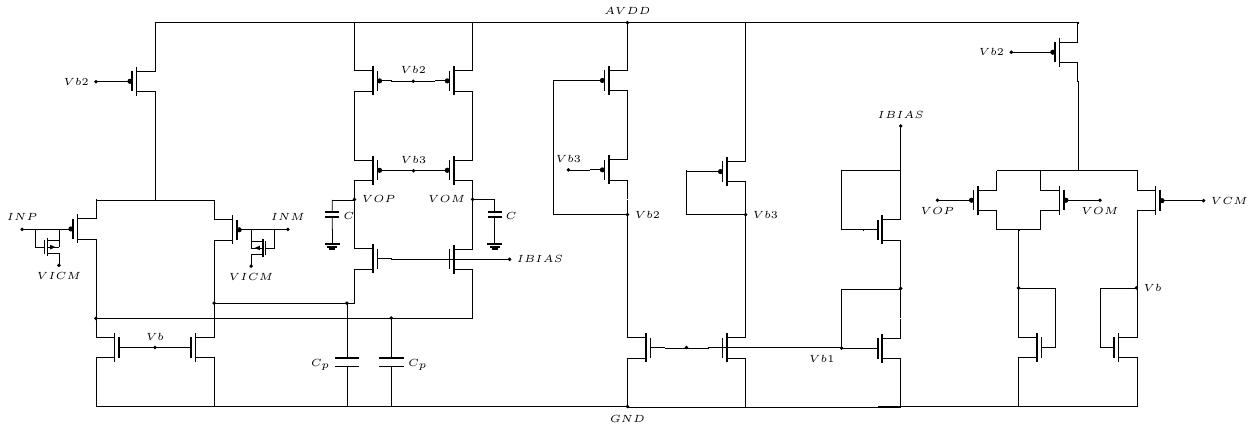}
    \caption{Folded cascode \gls{opamp/ota}.}
    \label{fig: fold_casc}
\end{figure*}

\begin{figure}[h!]
    \centering
    \includegraphics[width=\columnwidth]{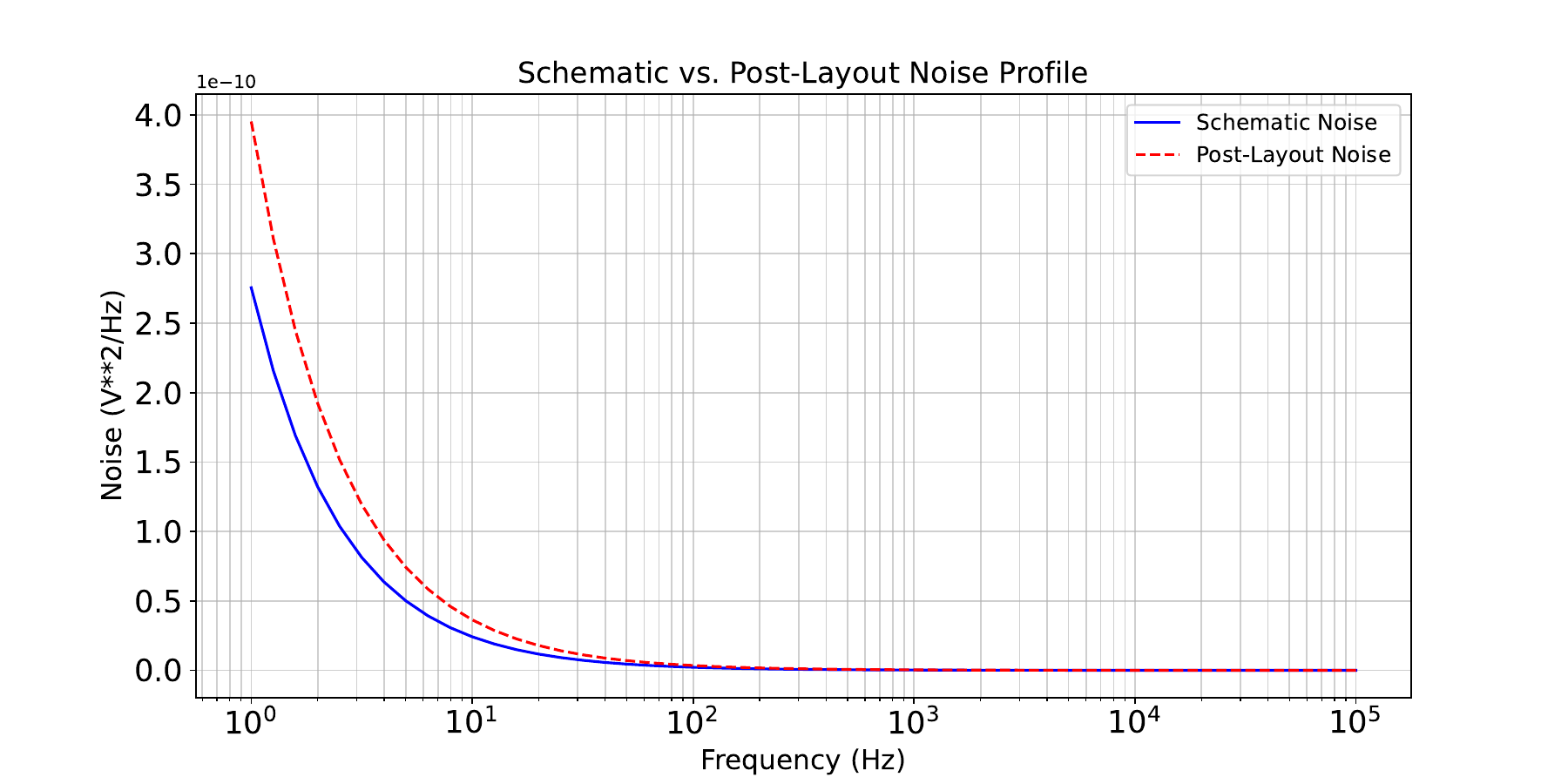}
    \caption{Noise profile of folded cascode \gls{opamp/ota}.}
    \label{fig: fold_casc_fc}
\end{figure}

The complete schematic of the folded cascode \gls{opamp/ota} is shown in  \cref{fig: fold_casc}. A low gain error amplifier is used for \gls{cmfb}. Together, with the high gain of the main folded cascode stage, the \gls{cmfb} loop gain is sufficiently high. Keeping the error amplifier gain low  helps in the \gls{cmfb} loop stability. More detailed description about the \gls{opamp/ota} is given in appendix: \cref{app: fold_casc}.
The value of \gls{fc} is around  1~KHz (see \cref{fig: fold_casc_fc}), hence, \gls{fch} is chosen to be 5~KHz.

\subsection{Results}
\begin{figure}
    \centering
    \includegraphics[width=\columnwidth]{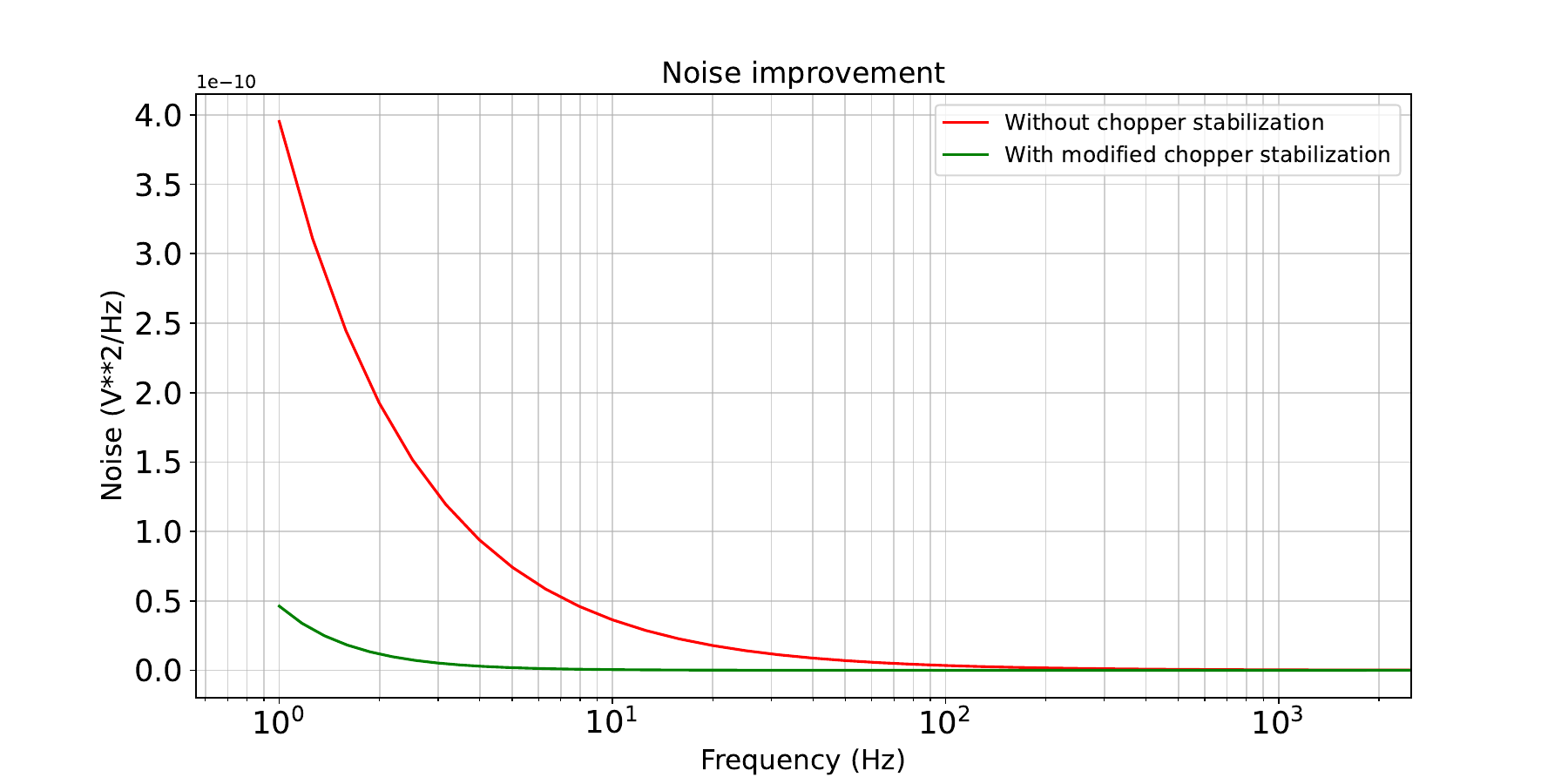}
    \caption{Proven noise improvement of this modulation scheme.}
    \label{fig: noise_imp}
\end{figure}

\begin{figure}[h!]
    \centering
    \includegraphics[width=\columnwidth]{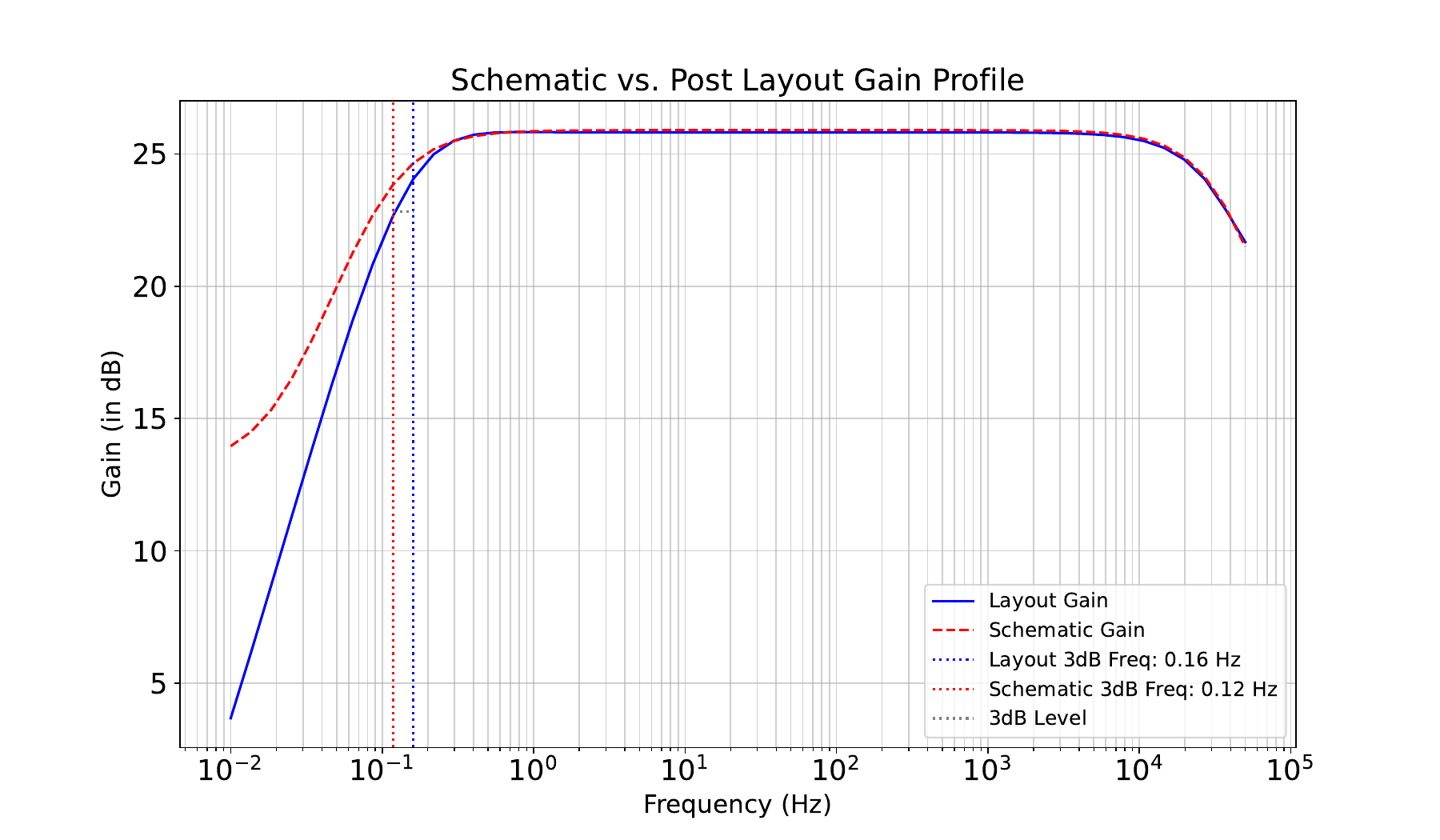}
    \caption{Gain comparison: from PAC analysis.}
    \label{fig: gain}
\end{figure}

\begin{figure}[h!]
    \centering
    \includegraphics[width=\columnwidth]{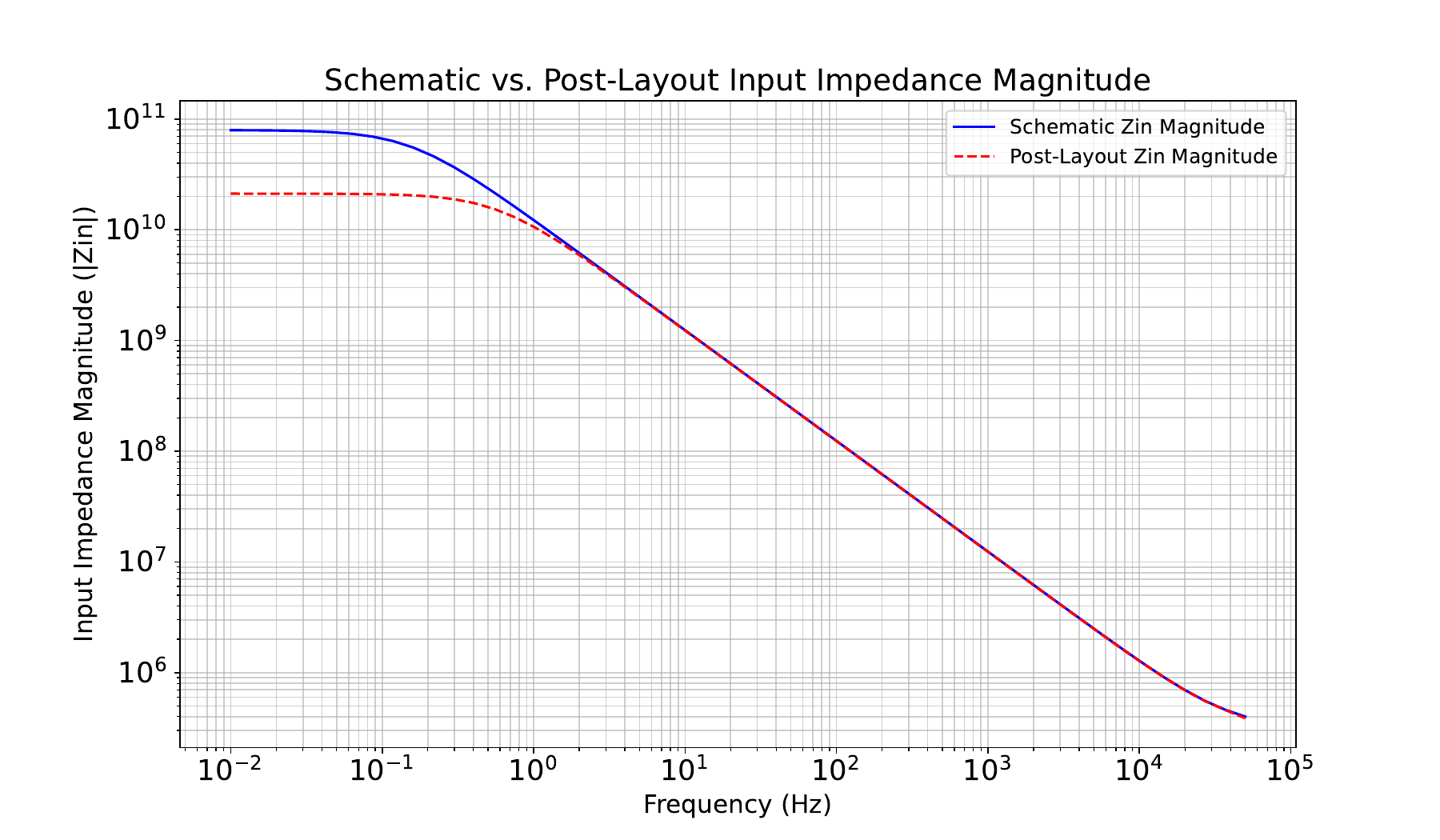}
    \caption{Input impedance: from PSS analysis.}
    \label{fig: zin}
\end{figure}

The closed loop gain plot is shown in  \cref{fig: gain}. The high pass corner is at 0.16~Hz and a mid-band gain of 26~dB is achieved.  \cref{fig: zin} shows the input impedance magnitude measures $>$~80~M$\Omega$ at 150~Hz and 21~G$\Omega$ at DC, which are suitable to interface with dry electrodes. The reported power consumption is 2.8~$\mu W$ with 2.6~$\mu W$ consumed by the folded cascode \gls{opamp/ota}. A total integrated noise of 7.2~$\mu$V\textsubscript{rms} is observed. \cref{fig: noise_imp} shows an improvement in noise profile after employing the proposed architecture.

\subsection{Inferences and Comparison with reported works}
The proposed architecture improves as technology scales. The DC \gls{Zin} increases with technology scaling, while the high pass corner frequency shifts toward the origin. Unlike earlier reported techniques for increasing \gls{Zin}~\cite{chandrakumar201727,pos_ccia_2}, the proposed technique does not require high gain for achieving the said objectives. The \gls{Zin} and high pass corner frequency improve as speed of digital gates and switches improves and parasitic capacitance reduce.

There is an implementation nuance to be considered while designing this circuit. The nodes $(D^+,D^-)$ are high impedance. Any asymmetric clock feed-through from $CH_3$ and $CH_4$ to these nodes can lead to offset. To prevent this, an extra capacitance can be deliberately added at the nodes $(D^+,D^-)$. An extra capacitance of 40~fF was added to the above nodes. This extra capacitance will also affect the high pass corner. The analytical results for the proposed circuit are shown in \cref{tab: ana_calc_pref} with parasitic values from post layout extraction.

\begin{table}[h!]
\renewcommand{\arraystretch}{2.5}
\caption{Analytical results}
\label{tab: ana_calc_pref} 
\centering
\begin{tabular}{|| c  | c ||}
\hline
Design parameter & Value \\
\hline
Input Impedance at DC  & 33~G$\Omega$ \\
\hline
Input Impedance at 150Hz  & 83~M$\Omega$ \\
\hline
Mid band Gain     & 26~dB    \\
\hline
High Pass Corner & 0.146~Hz \\  
\hline
\end{tabular}
\end{table}

The parameters of the design developed using this architecture are compared with the existing techniques in \cref{tab: perf}. All power numbers mentioned do not account for the power consumed by the clock generation circuit. 
\begin{table*}[t]
\caption{Performance Comparison}
\label{tab: perf}
\centering
\resizebox{\textwidth}{!}{%
\begin{tabular}{lcccccc}
\toprule
Method 
& This Work 
&  ~\cite{chandrakumar2017high} 
&  ~\cite{tuDelft2} 
&  ~\cite{china1}
&   ~\cite{pos_ccia_2}
&  ~\cite{ref_0} \\
\midrule

Technology (nm)                & 65   & 40 & 65 & 180 & 110 & 180  \\
V\textsubscript{DD} (V)        & 1.2  & 1.2 & 1 & 1.8 & 1.5 & 1.8\\
Power~($\mu$W)                 & 2.6  & 1.88 & 2.1 & 4.5 & 3.83 & 4  \\
$Z_\text{in}$ at DC            & 21~G$\Omega$ & 300~M$\Omega$ & 30~M$\Omega$ & $\ge$ 1~G$\Omega$ & 15~G$\Omega$ & NA  \\
DC servo loop required         & No  & Yes & Yes & Yes & Yes & Yes \\
$V_{\text{in,ref,noise}}$      & 7.2~$\mu V_{rms}$  & 2~$\mu V_{rms}$  & 0.7~$\mu V_{rms}$ & 0.2~$\mu V_{rms}$ & 0.54$\mu V_{rms}$& 0.73~$\mu V_{rms}$/3.5~$\mu V_{rms}$  \\
                               & (1~Hz-150~Hz) & (1~Hz-200~Hz) & (1~Hz-150~Hz)  & (3~Hz-300~Hz)  & (0.5~Hz-40~Hz)& (1-200~Hz/200-5K~Hz)\\
Dry electrode compatible       & Yes  & No & No & Yes & Yes & No  \\
Gain~(dB)                      & 26   & 26 & 40 & 60 & 64 & 41.5 \\
Bandwidth~(Hz)                 & 44~K   & 5~K & 0.5~K & 300 & 300 & 10.5~K  \\
CMRR at 50~Hz~(dB)             & 62   & NA & NA & 90 & 84 & NA  \\
PSRR at 50~Hz~(dB)             & 124   & NA & NA & 146 & 91(at 40~Hz) & NA  \\
High pass corner~(Hz)          & 0.16  & 0.2 - 4 & 0.5 & NA(\gls{dsl} not used) & 0.2 & 2.86 - 4\\
\bottomrule
\end{tabular}%
}
\end{table*}

\section{Conclusion}
Chopper stabilization commonly used in amplifiers for suppressing offset and \gls{1/f} noise suffers from low \gls{Zin}. Reported techniques of improving \gls{Zin} like pre-charge based techniques or positive feedback based techniques require high analog loop gain, strict matching constraints and are architecturally complex solutions. We report a new technique that leverages the charge stored on the differential capacitors in
the previous chopping phase to achieve high  \gls{Zin}. This technique only requires additional switches, and no additional analog amplifiers are needed. This technique removes the interdependency between \gls{fch} and \gls{Zin} allowing these two parameters to be chosen independently. 
The proposed circuit implicitly has a high-pass corner at a very low frequency making it suitable for acquisition of bio-potential signals like ECG/EMG. Example implementation in TSMC~65~nm CMOS features a  \gls{Zin} of 21~G$\Omega$ at DC with a bandwidth from 0.16~Hz to 150~Hz while consuming 2.8~$\mu$W of power. 


\appendices
\section{Modified \gls{cmfb} stage for folded cascode operating in sub threshold regime}
    \label{app: fold_casc}

    In the design shown in  \cref{fig: fold_casc}, load capacitors at the outputs create the dominant pole (not shown in the figure). Typically, the poles/zeros created by all other nodes should be 
    overshadowed by this dominant pole. But, the transistors here operate in or at the edge of sub-threshold region, due to which the nodes have high impedances. This reduces the frequencies of the non-dominant poles/zeros, compromising the stability of the \gls{cmfb} loop.
    To address this issue, the solution needs to be twofold:
    \begin{enumerate}
        \item Lower the loop gain.
        \item Position the poles/right half zeros generated by high impedance nodes beyond unity gain frequency.
    \end{enumerate}

    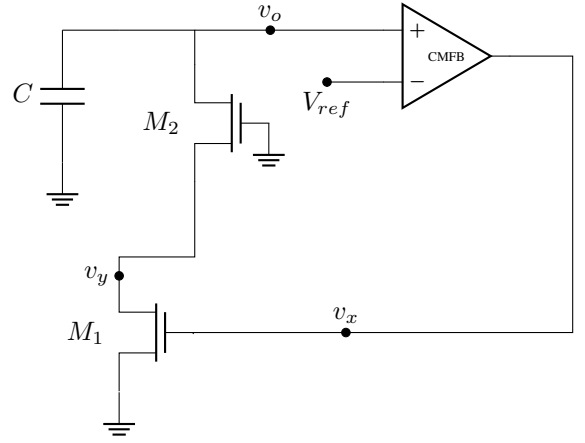
\begin{figure}[h!]
    \centering
 \begin{tikzpicture}
    \node[nmos, xscale=-1](N1) at (0, 3){} node[anchor=east] at (N1.text){$M_1$};
    \node[nmos, xscale=-1](N2) at (1, 5.77){} node[anchor=east] at (N2.text){$M_2$};
    \draw (0, 3.75) -| (0, 4) -- (1, 4) -| (1, 5);
    \node[ground] at (1.98, 5.77){};
    \node[ground] at (0, 2.21){};
    \draw (-0.75, 7) to[capacitor, /tikz/circuitikz/bipoles/length=0.980cm, l_={$C$}] (-0.75, 5.25);
    \node[ground] at (-0.75, 5.25){};
    \draw (0, 7) -- (1, 7) -- (1, 6.54);
    \draw (1, 7) -- (3.5, 7);
    \node[op amp, xscale=0.7, yscale=-0.7](N3) at (4.333, 6.657){} node[anchor=center] at (N3.text){\tiny CMFB};
    \draw (3.5, 6.314) -| (2.75, 6.25);
    \draw (5.166, 6.657) -| (6, 6.5) -| (6, 3) -| (0.98, 2.98);
    \node[circ](N4) at (2.75, 6.314){} node[anchor=north] at (N4.text){$V_{ref}$};
    \node[circ](N5) at (3, 3){} node[anchor=south] at (N5.text){$v_x$};
    \draw (0, 7) -- (-0.75, 7);
    \node[circ](N6) at (0, 3.75){} node[anchor=east] at (N6.text){$v_y$};
    \node[circ](N7) at (2, 7){} node[anchor=south] at (N7.text){$v_o$};
\end{tikzpicture}
    \caption{Common mode equivalent circuit.}
    \label{fig: eq_ckt_lpg}
    \end{figure}

    To reduce the loop gain consider the common mode equivalent circuit shown in  \cref{fig: eq_ckt_lpg}, where there are two amplifying stages. The cascode amplifier formed by M\textsubscript{1} and M\textsubscript{2} have very low drain currents and hence, a  high output impedance. The only way to reduce gain here would be to increase the current value, but that would alter all other parameters. So the gain of error amplifier can be lowered. In this design we have chosen a gain of unity.

\begin{figure}
        \centering
        \includegraphics[width=\columnwidth]{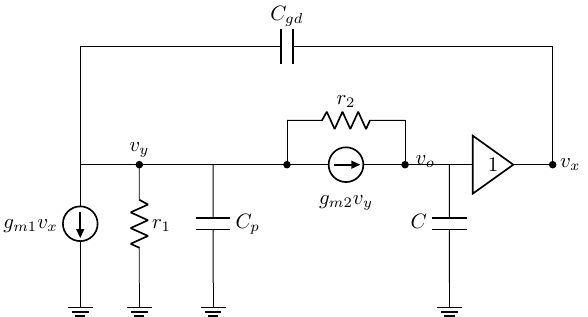}
        \caption{Small signal equivalent of common mode equivalent circuit.}
        \label{fig: ss_eq_ckt_lpg}
    \end{figure}

    To position the poles generated by high impedance nodes beyond unity gain frequency; again consider the common mode equivalent circuit shown in  \cref{fig: eq_ckt_lpg} with parasitic contributions from M\textsubscript{1} and M\textsubscript{2} as described below (the small signal model shown in  \cref{fig: ss_eq_ckt_lpg}).

    \begin{itemize}
        \item $C_{p}$: Parasitic capacitor at node $v_y$.
        \item $C_{gd}$: Parasitic capacitor between nodes $v_{y}$ and $v_x$.
        \item $g_{m1}$: Equivalent transconductacne of $M_1$.
        \item $g_{m2}$ : Equivalent transconductacne of $M_2$.
        \item $r_1$ : Equivalent resistance at drain of $M_1$.
        \item $r_2$: Equivalent resistance at drain of $M_2$.
    \end{itemize}

\begin{figure}[htbp]
    \centering
    \begin{minipage}{0.48\columnwidth}
        \centering
        \includegraphics[width=\linewidth]{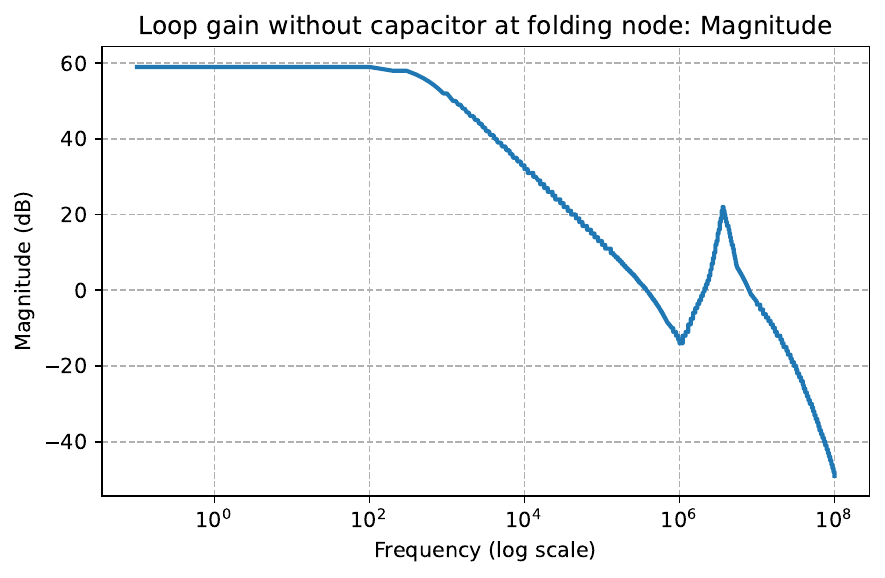}
    \end{minipage}
    \hfill
    \begin{minipage}{0.48\columnwidth}
        \centering
        \includegraphics[width=\linewidth]{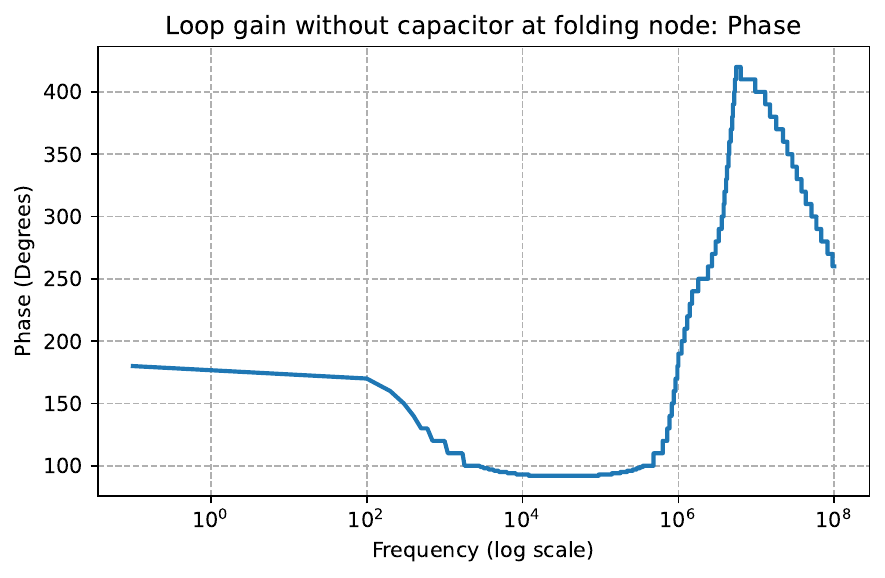}
    \end{minipage}
    \caption{Before adding capacitor at node $v_y$ of \cref{fig: ss_eq_ckt_lpg}.}
    \label{fig: lpg_b4_cp}
\end{figure}

\begin{figure}[htbp]
    \centering
    \begin{minipage}{0.48\columnwidth}
        \centering
        \includegraphics[width=\linewidth]{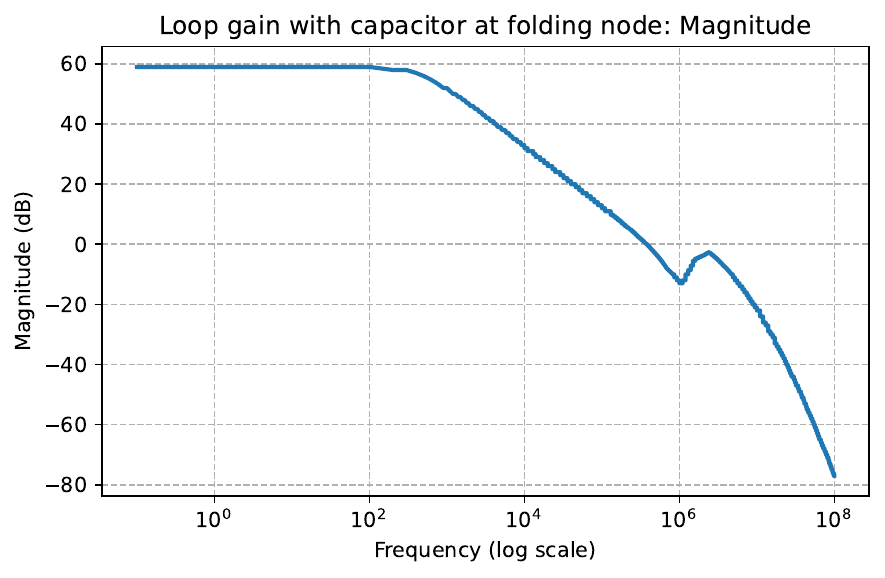}
    \end{minipage}
    \hfill
    \begin{minipage}{0.48\columnwidth}
        \centering
        \includegraphics[width=\linewidth]{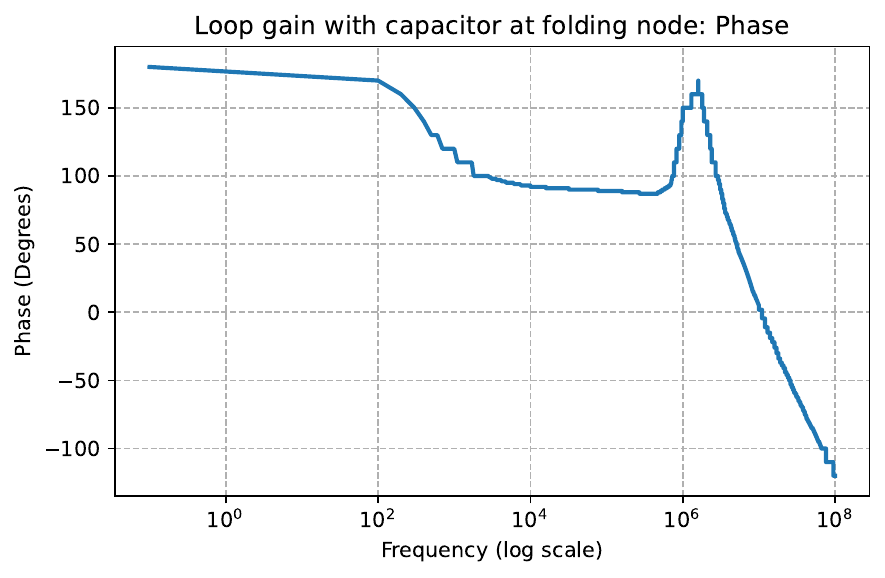}
    \end{minipage}
    \caption{After adding capacitor at node $v_y$ of \cref{fig: ss_eq_ckt_lpg}.}
    \label{fig: lpg_after_cp}
\end{figure}

    Writing KCL at nodes v\textsubscript{y} and v\textsubscript{o}:
\begin{align}
\begin{split}
&\frac{v_y}{r_1} + \frac{v_y - v_o}{r_2} + s C_p v_y + g_{m2} v_y \\
&\qquad + g_{m1} v_x + s C_{gd} (v_y - v_x) = 0
\end{split} \\[6pt]
& s C\, v_o + \frac{v_o - v_y}{r_2} - g_{m2} v_y = 0
\end{align}

\begin{equation}
\begin{split}
\frac{v_o}{v_x} =
\frac{r_{1}\left(C_{gd} g_{m2} r_{2} s + C_{gd}s
- g_{m1} g_{m2} r_{2} - g_{m1}\right)}
{ C C_{gd} r_{1} r_{2} s^{2}
+ C C_{p} r_{1} r_{2} s^{2}
+ C g_{m2} r_{1} r_{2} s  } \\
\qquad \times
\frac{1}{ C r_{1} s + C r_{2} s
+ C_{gd} r_{1} s + C_{p} r_{1} s + 1 }
\end{split}
\label{eq:  tf_eq}
\end{equation}
    This gives a zero at $Z_1 = \frac{g_{m1}}{c_{gd}}$, and two poles are apparent from denominator of~\cref{eq: tf_eq} which are cloddish to express here\footnote{The actual non-dominant poles are even more cloddish due to higher order system.}. 
    The dominant poles primarily depends on $C$ and output impedance of \gls{opamp/ota}. The first non-dominant pole depends on product of $C_p$ and $r_1$.
    The real problem is the right half zero which occurs before the non-dominant poles
    due to smaller $g_{m1}$ value. To counter this we can deliberately add some capacitance at node $v_{y}$ such that one of the non-dominant pole lies before this right half zero and after the \gls{ugb}. In this design, \gls{ugb} $\approx$ 1.5~MHz and r\textsubscript{1} $\approx$ 1M$\Omega$. We find that $C_p$ $\approx$ 100~fF is required for an  appropriate placement of the non-dominant pole. \Cref{fig: lpg_b4_cp} and~\cref{fig: lpg_after_cp}  show that the  phase margin is improved to $\approx$~\ang{80}. This is an unconventional way of stabilizing the  \gls{cmfb} loop. Usually node $v_x$ is the choice of compensation capacitor, but since it is a low impedance node the value of capacitor required would be impractical.~\footnote{To quantify, it will be in the order of nano-farads.}

\ifCLASSOPTIONcaptionsoff
  \newpage
\fi



%

\bibliographystyle{unsrt}
\bibliography{biblography}

%




\end{document}